\documentclass{jfm}

\usepackage{amsmath,amssymb}
\usepackage{graphicx}
\usepackage{color}

\newcommand\ii{{\rm{i}}}

\newcommand\bff{\textbf{f}}

\newcommand\bh{\textbf{h}}
\newcommand\bk{\textbf{k}}

\newcommand\bu{\textbf{u}}
\newcommand\bw{\textbf{w}}
\newcommand\bx{\textbf{x}}
\newcommand\bz{\textbf{z}}

\shorttitle{Inverse cascade of energy in helical turbulence}
\shortauthor{F. Plunian, A. Teimurazov, R. Stepanov and M.K. Verma}

\title{Inverse cascade of energy\\in helical turbulence}

\author{Franck Plunian\aff{1}
  \corresp{\email{Franck.Plunian@univ-grenoble-alpes.fr}},
  Andrei Teimurazov\aff{2},
 Rodion Stepanov\aff{2}
 \and 
 Mahendra Kumar Verma\aff{3}}

\affiliation{\aff{1}Universit\'e Grenoble Alpes, Universit\'e Savoie Mont Blanc, CNRS, IRD, IFSTTAR, ISTerre, 38000 Grenoble, France
\aff{2}Institute of Continuous Media Mechanics,  Korolyov 1,
Perm, 614013, Russia
\aff{3}Department of Physics, Indian Institute of Technology, Kanpur 208016, India}

\begin{document}

\maketitle

\begin{abstract}
Using direct numerical simulation of hydrodynamic turbulence with helicity forcing applied at all scales, a  near-maximum helical turbulent state is obtained, with an inverse energy cascade at scales larger than the energy forcing scale and a forward helicity cascade at scales smaller than the energy forcing scale. 
In contrast to previous studies using decimated triads, our simulations contain all possible triads. 
By computing the shell-to-shell energy fluxes, we show that the inverse energy cascade results from weakly non-local interactions among homochiral triads.
Varying the helicity injection range of scales leads to necessary conditions to obtain an inverse energy cascade. 
\end{abstract}

\maketitle
\section{Introduction}
%%%%%%%%%%
Inverse cascade of energy is a well-known feature of two-dimensional homogeneous isotropic turbulence (HIT). 
It is generally understood as a consequence of the positive-definiteness of two  ideally conserved quantities, energy and enstrophy, the enstrophy cascade being forward \citep{Kraichnan1967}. In three-dimensional HIT, in addition to energy, the other quantity conserved in the inviscid limit is helicity  \citep{Moreau1961,Moffatt1969}, which is not positive-definite. Accordingly only forward cascades of both energy and helicity are expected \citep{Brissaud1973,Chen2003,Chen2003b,Mininni2006,Alexakis2017}. However, projecting the velocity field on a basis of homochiral modes, e.g. positive helical modes, helicity thus becomes positive-definite, and an inverse cascade of energy is expected together with a forward cascade of helicity. This has been shown numerically by solving a decimated model of the Navier--Stokes equations (NSE) in which the negative helical modes have been arbitrarily set to zero \citep{Biferale2012,Biferale2013}. Limited to interactions among positive helical modes, the energy flux has been found to be negative at infrared scales, i.e. scales larger than the energy forcing scale, leading to an inverse cascade of energy. 

Although such a homochiral framework is theoretically interesting, it is not yet clear whether it could be applied to the NSE that involve both homochiral and heterochiral triadic interactions. 
\citet{Kessar2015} solved the NSE numerically, with energy injected at large scales and positive helicity injected over the entire inertial range of the energy cascade.
The energy of the positive helical modes was found to be higher than the energy of the negative helical modes by several orders of magnitude, corresponding to a near-maximum helical state of turbulence. They found that the energy flux resulting from the positive helical mode interactions was indeed negative, as predicted by the homochiral phenomenology. However, the energy flux resulting from the heterochiral interactions was found to be positive and dominant, leading to a positive total energy flux, and therefore to a forward energy cascade. Finally, the energy spectrum was found to satisfy a power law close to $k^{-7/3}$, related to the forward cascade of helicity in homochiral turbulence. 

In \citet{Sahoo2017}, the nonlinear operator of the NSE has been modified in order to control the relative weight of homochiral to heterochiral triadic interactions. 
In the high-Reynolds-number limit, by increasing the weight of homochiral triadic interactions, they found a sharp transition from forward to inverse energy cascade at infrared scales. It remains to be seen whether such an inverse cascade can occur dynamically from the direct numerical simulation of the NSE, without any type of decimation, which is the subject of this paper. 

We report results obtained from the direct simulation of the NSE, with positive helicity injected at all scales, as in \citet{Kessar2015}, but now including the scales larger than the energy forcing scale, namely the infrared scales. Although the flow is again not in an ideal state of maximum helicity because it contains both positive and negative helical modes interacting dynamically, an inverse cascade of energy is nevertheless found. This is the first evidence of inverse energy cascade in three-dimensional homogeneous isotropic turbulence, in a framework more general than that of homochiral triads.

\section{Helical forcing}
%%%%%%%%%%%%%%%%%%%
We solve the NSE
\begin{equation}
\label{eq:NS}
\partial_t\bu=-(\bu\cdot \nabla)\bu -\nabla P+ \nu \nabla^2\bu - \mu \nabla^{-4}\bu +\bff,
\end{equation}
where $\bu$ denotes the velocity field, $P$ the pressure normalized by the mass density, $\nu$ the fluid viscosity and $\bff$ the flow forcing.
We consider an incompressible fluid such that $\nabla \cdot \bu = 0$.
We introduce the additional term $\mu \nabla^{-4}\bu$, which mimics large-scale friction in order to avoid large-scale energy accumulation in case an inverse cascade occurs.

In Fourier space, equation (\ref{eq:NS}) becomes
\begin{equation}
\label{eq:NSFourier}
\partial_t \bu(\bk)=-\mathcal{F}[(\bu\cdot\nabla)\bu](\bk) -\ii P(\bk) \bk
-(\nu k^2 +\mu k^{-4})\bu(\bk) +\bff(\bk),
\end{equation}
where $\mathcal{F}[(\bu\cdot\nabla)\bu](\bk)$ denotes the Fourier transform of the nonlinear term $(\bu\cdot\nabla)\bu$, and $k=|\bk|$.
In (\ref{eq:NSFourier}) $\bu(\bk)$, $P(\bk)$ and $\bff(\bk)$ denote the Fourier coefficients of, respectively, $\bu$, $P$ and $\bff$ at wavenumber $\bk$. The same notation will be used throughout the paper.

A crucial issue is to derive a forcing $\bff(\bk)$ such that helicity $H(\bk)=\frac{1}{2}\bu(\bk)\cdot\bw(\bk)^*$, where $\bw(\bk)=i \bk \times \bu(\bk)$ is the vorticity, can be injected independently from energy $E(\bk)=\frac{1}{2}\bu(\bk)\cdot\bu(\bk)^*$. Therefore $\bff(\bk)$ has to satisfy
\begin{eqnarray} \label{eq:forcing}
\bff(\bk)\cdot \bu(\bk)^*= \varepsilon_E(\bk), \;\;\;\;\;\;
\bff(\bk)\cdot \bw(\bk)^*= \varepsilon_H(\bk),
\end{eqnarray}
where $\varepsilon_E(\bk)$ and $\varepsilon_H(\bk)$ are the injection rates of, respectively, energy and helicity.

\subsection{Helical mode decomposition}
%%%%%%%%%%%%%%%%%%%
Each velocity Fourier coefficient is split into helical modes \citep{Craya1958,Herring1974,Cambon1989,Waleffe1992,Lessinnes2011}, 
\begin{equation}\label{eq:uhelmodes}
    \bu(\bk) =  u^+(\bk)\bh^+(\bk)+u^-(\bk)\bh^-(\bk) \equiv \bu^+(\bk)+\bu^-(\bk), 
\end{equation}
where $u^{\pm}(\bk)$ are complex scalars and $\bh^{\pm}(\bk)$ are the two eigenvectors of the curl operator, satisfying $\ii \bk\times\bh^{\pm}(\bk)=\pm k\bh^{\pm}(\bk)$.
The latter are defined as 
\begin{equation}\label{eq:helh}
 \bh^\pm(\bk)=\frac{1}{\sqrt{2}}\frac{(\bz_\bk\times \bk)\times \bk}{k|\bz_\bk\times \bk|}\pm\frac{\ii}{\sqrt{2}}\frac{\bz_\bk\times \bk}{|\bz_\bk\times \bk|},
\end{equation}
where at each time step the vector $\bz_\bk$ is generated randomly for each wavenumber $\bk$, keeping $\bz_\bk \times \bk \neq 0$ \citep{Waleffe1992}. 
Note that $\bh^+(\bk)$ and $\bh^-(\bk)$ are complex conjugate of each other, $\bh^\pm(\bk)=\bh^\mp(\bk)^*$.  They are vectors of unit norm, $|\bh^\pm(\bk)|^2=\bh^+(\bk)\cdot \bh^-(\bk)=1$, and the scalar product of each of them with itself is equal to zero, $\bh^+(\bk)\cdot\bh^+(\bk)=\bh^-(\bk)\cdot\bh^-(\bk)=0$. Then the helical modes $u^\pm(\bk)$ can be derived from (\ref{eq:uhelmodes}) according to
\begin{equation}\label{eq:helf}
 u^\pm(\bk)=\bu(\bk)\cdot\bh^\mp(\bk).
\end{equation}
The vorticity can also be expressed in terms of helical modes,
\begin{equation}\label{eq:whelmodes}
    \bw(\bk)= k\left(u^+(\bk)\bh^+(\bk)-u^-(\bk)\bh^-(\bk)\right).
\end{equation}
Then, the energy and helicity take the form 
\begin{eqnarray}\label{eq:energy}
E(\bk)=E^+(\bk)+E^-(\bk), \;\;\;\;\;
H(\bk)=k\left(E^+(\bk)-E^-(\bk)\right),\label{eq:helicity2}
\end{eqnarray}
with $E^\pm(\bk)=\frac{1}{2}|u^\pm(\bk)|^2$.

\subsection{Dynamical forcing}
%%%%%%%%%%%%%%
Looking for a flow forcing of the form
\begin{equation}\label{eq:held}
 \bff(\bk)=c^+(\bk) u^+(\bk)\bh^+(\bk) + c^-(\bk) u^-(\bk)\bh^-(\bk),
\end{equation}
from (\ref{eq:forcing}), (\ref{eq:uhelmodes}) and (\ref{eq:whelmodes}), we obtain
\begin{equation}
c^\pm(\bk)=\frac{\varepsilon_E(\bk) \pm \varepsilon_H(\bk)/k}{4E^\pm(\bk)}. 
\end{equation}
As our purpose is to study strongly helical turbulence, one type of helical mode, here $\bu^-(\bk)$, is expected to have a small energy $E^-(\bk)$, possibly leading to high values of $c^-(\bk)$. To avoid using a very small time step, which would be numerically intractable, we integrate analytically the equations $ \partial u^\pm(\bk,t)/\partial t=c^\pm(\bk,t) u^\pm(\bk,t)$, leading to 
\begin{equation}\label{eq:sol1}
 u^\pm_0(\bk,t+\Delta t)=
 e^{\ii \arg{u^\pm(\bk,t)}}
\left(\left[\varepsilon_E(\bk,t) \pm \frac{\varepsilon_H(\bk,t)}{k}\right]\Delta t 
+|u^\pm(\bk,t)|^2\right)^{1/2}.
\end{equation}
Then $\bu_0(\bk,t+\Delta t)= u_0^+(\bk,t+\Delta t)\bh^+(\bk)+u_0^-(\bk,t+\Delta t)\bh^-(\bk)$
is taken as an initial condition in the calculation of the NSE at time step $t+\Delta t$. To achieve the desired state of high helicity, we use the following asymptotic quenching:
\begin{equation}
\varepsilon_H(\bk,t)=\tilde{\varepsilon}_H \left(1-\frac{|E^+(\bk,t) - E^-(\bk,t)|}{E^+(\bk,t) + E^-(\bk,t)}\right),
\end{equation}
where $\tilde{\varepsilon}_H$ is the strength of helicity injection rate.

\section{Energy and helicity}
%%%%%%%%%
\subsection{Integral quantities}
%%%%%%%%%%%%%%%%
The simulations are performed with the pseudo-spectral code TARANG \citep{Verma2013,Stepanov2017,Teimurazov2018}, in a triply periodic domain of size $(2\pi)^3$.
Energy is injected at the permanent rate $\varepsilon_E(\bk)=0.2$, over $|\bk|=k_E\in ]9,10]$. 
The viscosity and large-scale friction values are taken as $\nu = 1.5\times10^{-3}$ and $\mu = 5\times10^{-2}$, for which a resolution of $512^3$ is sufficient. 
The simulations differ by the values chosen for $\tilde{\varepsilon}_H$, which is applied over $|\bk|=k_H\in[1,10^2]$. 
Some variations of the parameters $\nu$ and $k_H$ will be discussed in section \ref{sec:varhel}.

\begin{figure}
	\centering
	\includegraphics[scale=0.65]{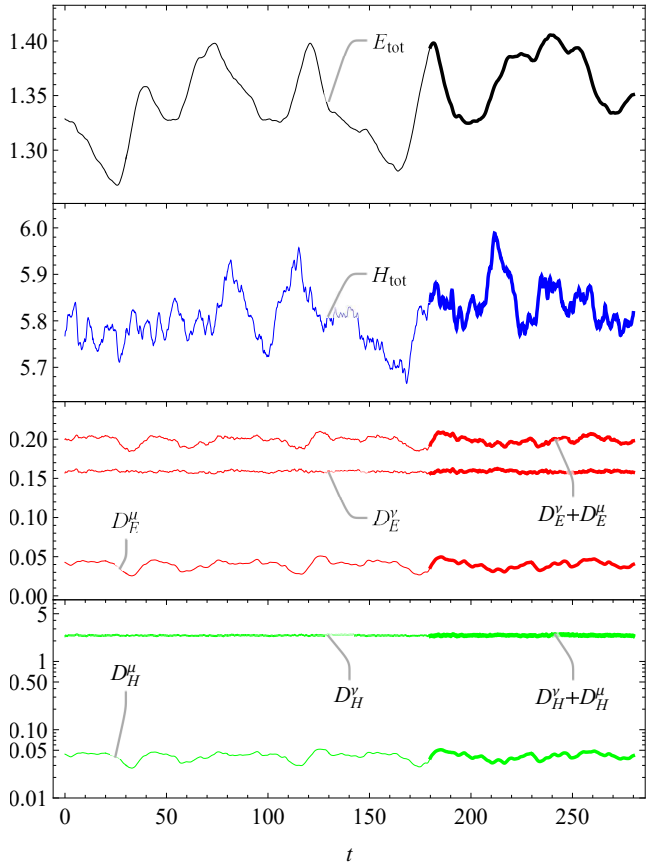}
	\caption{Total energy $E_{\rm tot}$, total helicity $H_{\rm tot}$, energy dissipation rates $D_E^\nu, D_E^\mu$ and $D_E^\nu+D_E^\mu$, helicity dissipation rates $D_H^\nu,D_H^\mu$ and $D_H^\nu+D_H^\mu$, are plotted versus time for $\varepsilon_E=0.2$ and $\tilde{\varepsilon}_H=25$. The time range $t\in[180,280]$ over which the statistics are calculated is indicated, for each curve, by a thicker line.
	}
	\label{fig:k10_Hk}
\end{figure} 

In Fig.~\ref{fig:k10_Hk}, several integral quantities are plotted versus time for $\tilde{\varepsilon}_H=25$. We define $X_{\rm tot}=\int X(\bk) d\bk$, with $X=E$ for total energy  and  $X=H$ for total helicity. The molecular dissipation rate of $X$ is given by $D_X^{\nu}=2\nu \int k^2X(\bk) d\bk$,  and the friction dissipation rate by $D_X^{\mu}=2\mu \int k^{-4}X(\bk) d\bk$.
In Fig.~\ref{fig:k10_Hk} we see that $H_{\rm tot}$ varies on a time scale much smaller than $E_{\rm tot}$, suggesting a turbulence governed by helicity rather than energy.  On average, we have   $D_E^{\nu}+D_E^{\mu} = 0.2$, which corresponds to the total injection rate of energy $\varepsilon_E=\int \varepsilon_E(\bk) d\bk$.
Both $D_E^{\mu}$ and $D_H^{\mu}$ are mostly identical. This is because friction is dominated by the largest scale $k=1$ where turbulence is close to maximum helicity, implying $H(k=1)\approx k E(k=1)$.
In the rest of the paper, the spectra and fluxes are obtained by averaging over frames spread over 50 to 100 time units. For $\tilde{\varepsilon}_H=25$, this time window is represented in Fig.~\ref{fig:k10_Hk} by the thick part of the curves.
  
An estimate of the Taylor-microscale Reynolds number, $R_{\lambda}=u'\lambda/\nu$, can be obtained with
$u'=\sqrt{2\bar{E}_{tot}/3}$ and  $\lambda=\sqrt{15 \nu u'^2/\varepsilon_E}$ \citep{Pope2000}.
From Fig.~\ref{fig:k10_Hk} the mean kinetic energy is $\bar{E}_{tot}\approx 1.35$, which gives 
$u'\approx 0.95$ and $\lambda \approx 0.32$,
thus leading to 
$R_{\lambda}\approx 200$, which compares well with other numerical simulations of the same resolution \citep{Okamoto2007}.

\subsection{Energy and helicity spectra and fluxes}
\label{sec:Energy and helicity spectra and fluxes}
%%%%%%%%%%%%%%%%%%
Taking the dot product of (\ref{eq:NSFourier}) with 
$\bx(\bk)^*$, we obtain the equations for modal energy and helicity,
\begin{eqnarray}
\label{eq:EFourier}
\partial_t X(\bk)=&-&\Re\{\bx(\bk)^*\cdot\mathcal{F}[(\bu\cdot\nabla)\bu](\bk)\} \nonumber\\
&-&2(\nu k^2 +\mu k^{-4})X(\bk) +\varepsilon_X(\bk),
\end{eqnarray}
where $(X,\bx)=(E, \bu)$ or $(X,\bx)=(H, \bw)$, $\mathcal{F}$ standing for Fourier transform, and $\varepsilon_X(\bk)$ satisfying (\ref{eq:forcing}).
Rewriting (\ref{eq:EFourier}) for $\bk'$, and taking the sum over $\bk'$ such that $k<|\bk'|\le k+dk$ \citep{Verma2019},
leads to 
\begin{equation}
\partial_t X(k)=-\partial_k\Pi_X(k)-D_X(k)+\varepsilon_X(k),
\label{eq:specequation}
\end{equation}
with
\begin{eqnarray}
X(k) dk&=&\sum_{k<|\bk'|\le k+dk} X(\bk'), \label{eq:specenergy}\\
\Pi_X(k)&=&\sum_{|\bk'|\le k} \Re\{\bx(\bk')^*\cdot\mathcal{F}[(\bu\cdot\nabla)\bu](\bk')\}, \label{eq:Fluxes}\\
D_X(k) dk&=&\sum_{k<|\bk'|\le k+dk} 2\left(\nu {k'}^2+\mu{k'}^{-4}\right) X(\bk') ,\\
\varepsilon_X(k) dk&=&\sum_{k<|\bk'|\le k+dk} \varepsilon_X(\bk'). \label{eq:specforcing}
\end{eqnarray}

\begin{figure}
	\centering
	\includegraphics[scale=0.75]{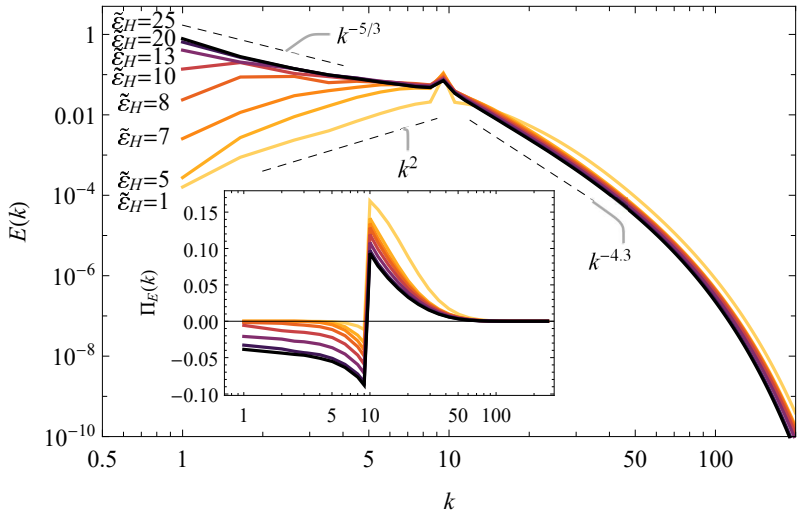}
	\caption{Energy spectra and fluxes (inlet). The energy injection rate $\varepsilon_E=0.2$ is applied at $k_E\in]9,10]$. The helicity injection rate  $\tilde{\varepsilon}_H\in \{1;5;7;8;10;13;20;25\}$ is applied at $k_H\in[1,10^2]$.}
	\label{fig:2}
\end{figure}

\begin{figure}
	\centering
	\includegraphics[scale=0.75]{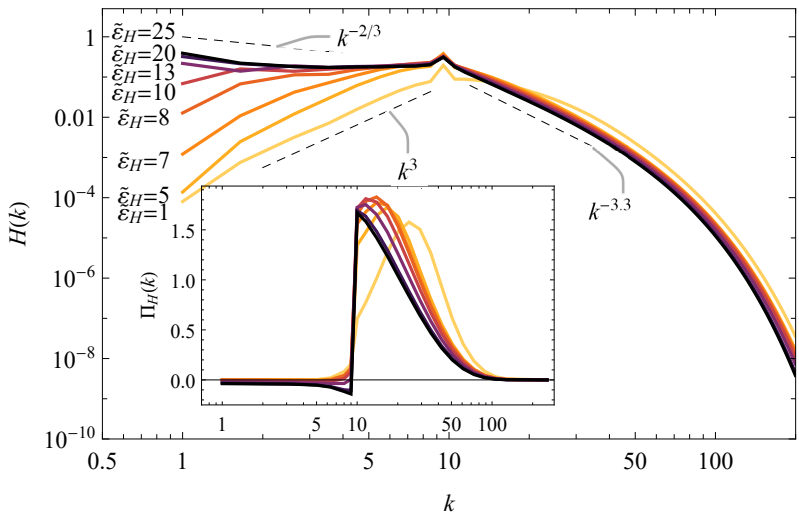}
	\caption{Helicity spectra and fluxes (inlet) for the same parameters as Fig.~\ref{fig:2}.}
	\label{fig:H}
\end{figure}

The energy spectral density $E(k)$ and flux $\Pi_E(k)$ in Fig.~\ref{fig:2}, and the helicity spectral density $H(k)$ and flux $\Pi_H(k)$ in Fig.~\ref{fig:H}, are plotted for $\tilde{\varepsilon}_H\in S$, with $S=\{1;5;7; 8; 10; 13; 20; 25\}$.
For $\tilde{\varepsilon}_H=1$, the fluxes of energy and helicity are non-zero only for $k\ge10$. As both fluxes are positive, they correspond to forward cascades of energy and helicity, as expected in three-dimensional turbulence.  In Fig.~\ref{fig:H}, the shape of the helicity flux obtained for $\tilde{\varepsilon}_H=1$, with a maximum at $k\approx 20$, is similar to the one obtained by \citet{Kessar2015} in which helicity was injected only at scales smaller than the energy injection scale.

With the increase of $\tilde{\varepsilon}_H$ from 1 to 25, the energy flux is shifted downwards.
It then becomes negative in the infrared domain $k\le 9$ , leading to an inverse cascade of energy.
With the increase of $\tilde{\varepsilon}_H$, the scale at which the helicity flux is maximum is also shifted, towards the energy injection scale.
For $k \ge10$ the helicity flux is positive, corresponding to a forward cascade of helicity. Such a dual cascade is consistent with the homochiral phenomenology.
For $\tilde{\varepsilon}_H=25$, in the infrared domain $k \le 9$,  the energy spectrum obeys a $k^{-5/3}$ scaling law. For $k\ge10$ the energy spectral slope is approximately $E(k)\propto k^{-4.3}$, which is steeper than the $E(k)\propto k^{-7/3}$ predicted by the homochiral phenomenology. Most probably this is due to the fact that, at scales smaller than the energy injection scales, the viscous dissipation cannot be neglected, implying that an inertial range can hardly be identified. 

\begin{figure}
	\centering
	\includegraphics[scale=0.75]{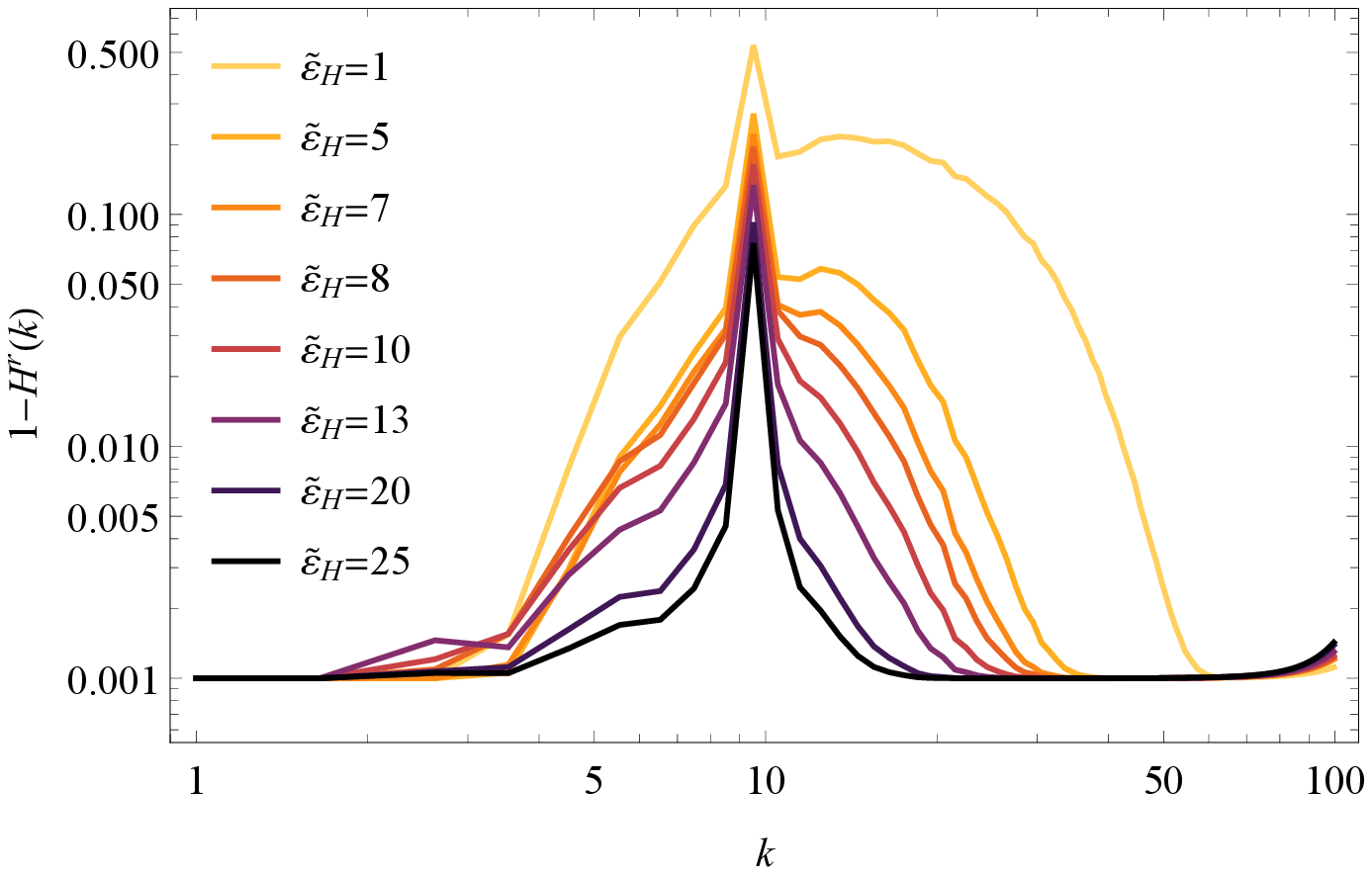}
	\caption{Deviation to maximum chirality,  $1-H^r(k)$, for the same parameters as Fig.~\ref{fig:2}.}
	\label{fig:Hr}
\end{figure}
A measure of chirality is given by the relative helicity $H^r(k)=H(k)/kE(k)$, which satisfies the realizability condition $0\le |H^r(k)| \le 1$. 
A non-helical turbulence corresponds to $|H^r(k)|=0$. A maximum helical turbulence corresponds to $|H^r(k)|=1$, which can be obtained only in the homochiral framework. 
In Fig.~\ref{fig:Hr}, $1-H^r(k)$ is plotted for $\tilde{\varepsilon}_H\in S$.
We find that increasing $\tilde{\varepsilon}_H$ leads to lower values of $1-H^r(k)$ in broader ranges of scales. The turbulent state is then closer to maximum chirality. We note that the lower bound of $10^{-3}$ for $1-H^r(k)$ is prescribed by a restriction on $\varepsilon_H$ in order to avoid a negative value under the square root in (\ref{eq:sol1}).

\section{Helical modes}
%%%%%%%%%%%%%%%%%%%%%%%%%%%%
\subsection{Energy fluxes}
%%%%%%%%%%%%%
The energy equation for each helical mode is obtained by taking the real part of the dot product of (\ref{eq:NSFourier}) with
$\bu^\pm(\bk)^*$. Then the energy equation is given by (\ref{eq:EFourier}) with $(X,\bx)=(E^\pm, \bu^\pm)$, and $2\varepsilon_{E^\pm}(\bk)=\varepsilon_E(\bk) \pm \varepsilon_H(\bk)/k$.
The equation satisfied by the energy spectral density of each helical mode is given by (\ref{eq:specequation}) with (\ref{eq:specenergy}-\ref{eq:specforcing}). Of course,  the flux of total energy satisfies
\begin{equation}
\Pi_E(k)=\Pi_{E^+}(k)+\Pi_{E^-}(k).
\label{eq:total flux}
\end{equation}
In addition,
each flux $\Pi_{E^\pm}(k)$ can be decomposed into
\begin{eqnarray}
\Pi_{E^+}(k)&=&\Pi^{+<}_+(k)+\Pi^{+<}_-(k),\\
\Pi_{E^-}(k)&=&\Pi^{-<}_+(k)+\Pi^{-<}_-(k),
\end{eqnarray}
where $\Pi^{a<}_{b}(k)$, with $a,b\equiv\pm$, denotes the energy flux from $\bu^{a<}(\bk)$, meaning that $\bu^{a}(\bk)$ is taken at wavenumbers inside a sphere
of radius $k$ (the $k$-sphere), to $\bu^{b}(\bk)$ taken at all wavenumbers. It is defined by
\begin{equation}
\label{eq:Fluxab}
    \Pi^{a<}_b(k)=\sum_{|\bk'|\le k} \Re\{\bu^a(\bk')^*\cdot \mathcal{F}[(\bu\cdot\nabla)\bu^b](\bk')\},
\end{equation}
where $\bu^b= \mathcal{F}^{-1}[\bu^b(\bk)]$ is the inverse Fourier transform of $\bu^b(\bk)$  \citep{Verma2004,Kessar2015,Plunian2019,Sadhukhan2019}.
We can further decompose $\Pi^{+<}_+(k)$ as $\Pi^{+<}_+(k)=\Pi^{+<}_{+<}(k)+\Pi^{+<}_{+>}(k)$, where $\Pi^{+<}_{+<}(k)$ denotes the energy flux from $\bu^{+<}$ to $\bu^{+<}$, and $\Pi^{+<}_{+>}(k)$ the energy flux from $\bu^{+<}$ to $\bu^{+>}$. Here, again, $\bu^{+<}$ and $\bu^{+>}$ denote  $\bu^+(\bk)$ taken at wavenumbers, respectively, inside and outside the $k$-sphere (similar notation applies to $\bu^{-<}$ and $\bu^{->}$).
By definition $\Pi^{+<}_{+<}(k)=0$, implying that $\Pi^{+<}_+(k)=\Pi^{+<}_{+>}(k)$ (similarly $\Pi^{-<}_{-<}(k)=0$ implies $\Pi^{-<}_-(k)=\Pi^{-<}_{->}(k)$).

\begin{figure}
	\centering
	\includegraphics[scale=0.2]{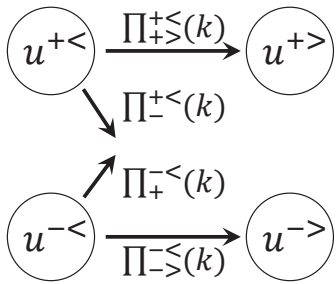}
	\caption{Illustration of the fluxes among helical modes. The notations $\bu^{\pm<}$ and $\bu^{\pm>}$ correspond to $\bu^\pm(\bk)$ taken at wave numbers respectively inside and outside the $k$-sphere. The non-horizontal arrows denote energy fluxes from one helical mode taken inside the $k$-sphere towards the opposite helical mode taken at all scales.}
	\label{fig:dessin}
\end{figure}

\begin{figure}
	\centering
	\includegraphics[scale=0.47]{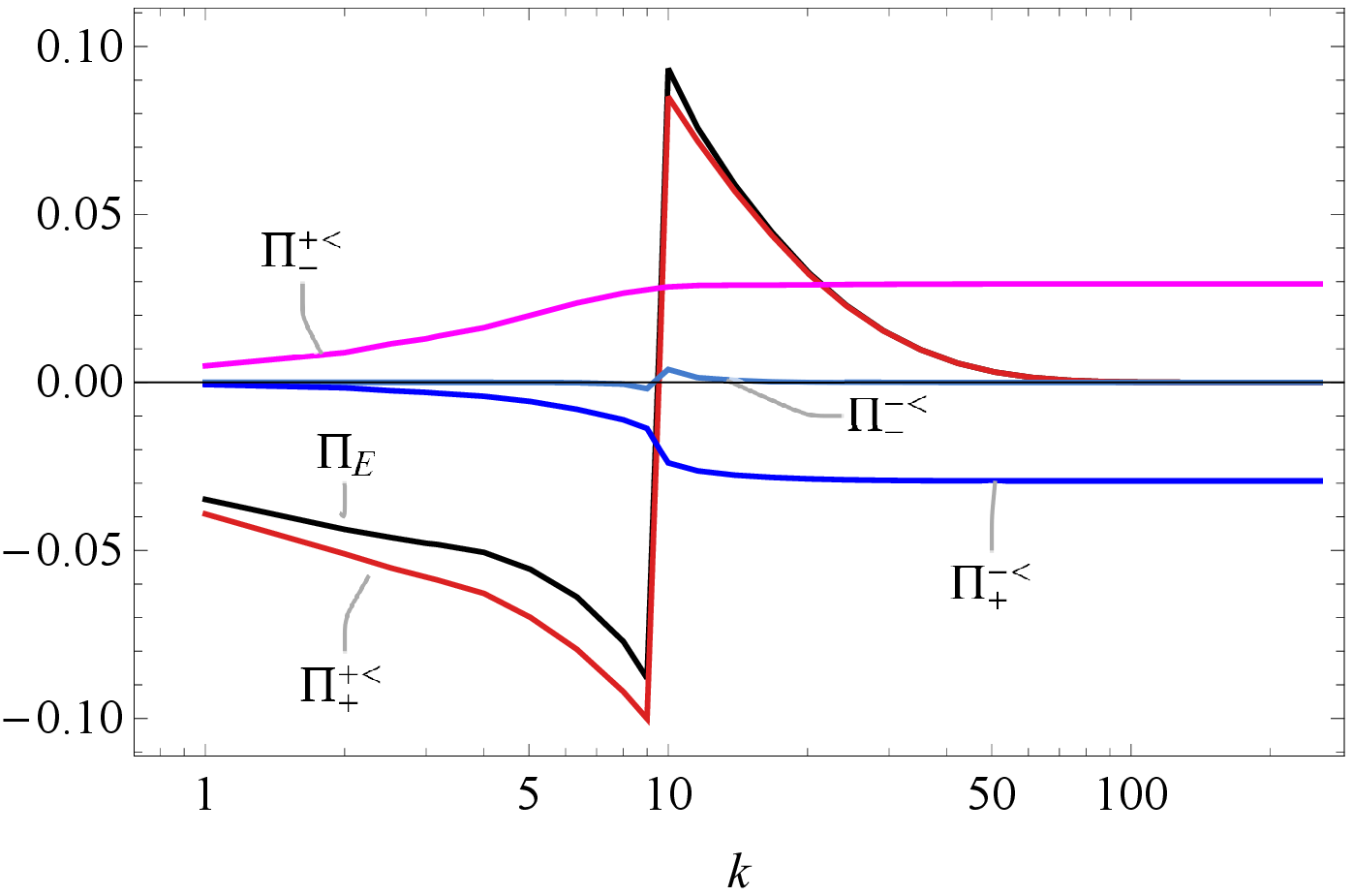}
	\includegraphics[scale=0.47]{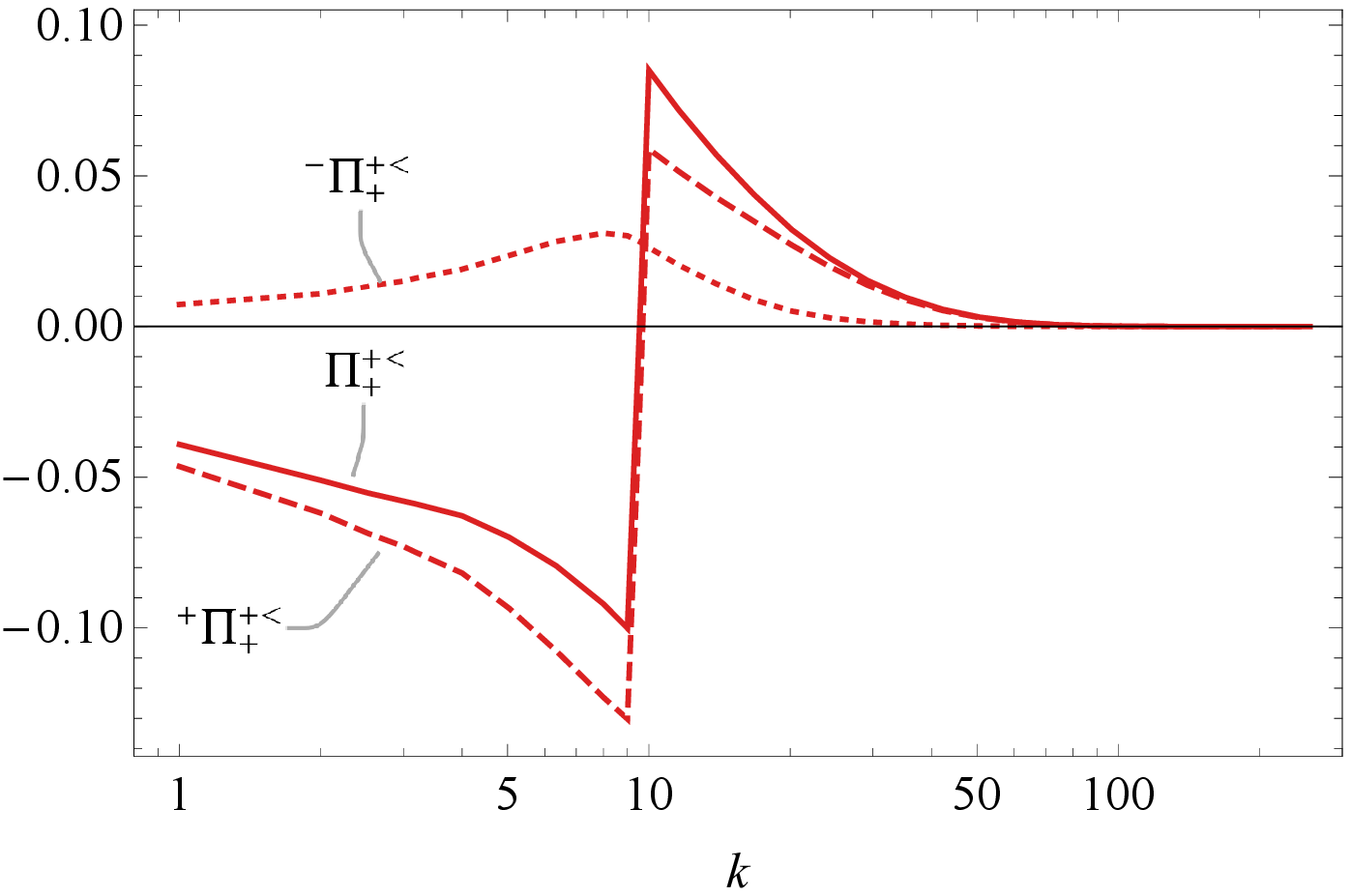}
	\caption{Helical mode fluxes for $\varepsilon_E=0.2$ and $\tilde{\varepsilon}_H=25$.  In the left figure, the four fluxes $\Pi^{+<}_{+}(k), \Pi^{+<}_-(k), \Pi^{-<}_+(k)$ and $\Pi^{-<}_{-}(k)$, illustrated in Fig.~\ref{fig:dessin}, are plotted together with the total energy flux $\Pi_E(k)$ already plotted in Fig.~\ref{fig:2}.
	In the right figure, the flux $\Pi^{+<}_{+}(k) $ is split into the homochiral flux ${}^+\Pi^{+<}_{+}(k)$ and the heterochiral flux ${}^-\Pi^{+<}_{+}(k)$.}
	\label{fig:5}
\end{figure}

An illustration of the four fluxes $\Pi^{+<}_{+}(k), \Pi^{+<}_-(k), \Pi^{-<}_+(k)$ and $\Pi^{-<}_{-}(k)$ is given in Fig.~\ref{fig:dessin}.
They are plotted in the left figure of Fig.~\ref{fig:5} for $\tilde{\varepsilon}_H=25$, together with $\Pi_E(k)$, which is equal to the sum of the four. In the infrared domain $k\in[1,9]$, $\Pi^{+<}_{+}(k)$ is clearly negative and close to $\Pi_E(k)$, suggesting that it is mostly this flux that is responsible for the inverse energy cascade.

It is thus interesting to further decompose the flux $\Pi^{+<}_{+}(k)$ as $\Pi^{+<}_{+}(k)={}^+\Pi^{+<}_{+}(k)+{}^-\Pi^{+<}_{+}(k)$
with
\begin{eqnarray}
{}^{\pm}\Pi^{+<}_{+}(k)
=\sum_{|\bk'|\le k}\Re\{\bu^+(\bk')^*\cdot {\cal{F}}[(\bu^{\pm}\cdot\nabla)\bu^+](\bk')\},
\label{flux2}
\end{eqnarray}
where ${}^{\pm}\Pi^{+<}_{+}(k)$ is interpreted as the energy flux from $\bu^{+<}$ to $\bu^{+}$, with $\bu^{\pm}$ acting as a mediator.

The three fluxes $\Pi^{+<}_+(k), {}^{+}\Pi^{+<}_{+}(k)$ and ${}^{-}\Pi^{+<}_{+}(k)$ are plotted in the right figure of Fig.~\ref{fig:5}. 
In the infrared domain $k\in[1,9]$, ${}^{+}\Pi^{+<}_{+}(k)\le 0$ and ${}^{-}\Pi^{+<}_{+}(k) \ge 0$, suggesting that the homochiral triadic interactions $(+,+,+)$ are mostly responsible for the inverse energy cascade. At scales smaller than the energy injection scales, $k \ge 10$,
we observe that ${}^{+}\Pi^{+<}_{+}(k)\ge 0$, which is in striking difference with previous simulations in which energy was injected at large scales \citep{Kessar2015}. In addition $|{}^{+}\Pi^{+<}_{+}(k)|\ge |{}^{-}\Pi^{+<}_{+}(k)|$, which contradicts the assumptions of the phenomenological model by \citet{2015PhRvL}.

From Fig.~\ref{fig:2} we observe that the inverse cascade is continuously greater for increasing values of $\tilde{\varepsilon}_H$. This is in contrast with 
 \citet{Sahoo2017} in which a sharp transition between the cases with and without inverse cascade is reported. There are at least two main reasons to explain this discrepancy.
First, in \citet{Sahoo2017} such a sharp transition is only expected within the limit of high Reynolds numbers, which we are not studying here. Second, in \citet{Sahoo2017}, the ratio of heterochiral to homochiral triadic interactions is imposed, and it is imposed in the same way at all scales. This was done by solving a modified version of the NSE including a prescribed weighting between the homochiral and heterochiral nonlinear terms. 
Here, as we solve the NSE (\ref{eq:NSFourier}), the importance of homochiral versus heterochiral triadic interactions is not prescribed, and remains scale-dependent.

\begin{figure}
	\centering
	\includegraphics[scale=0.7]{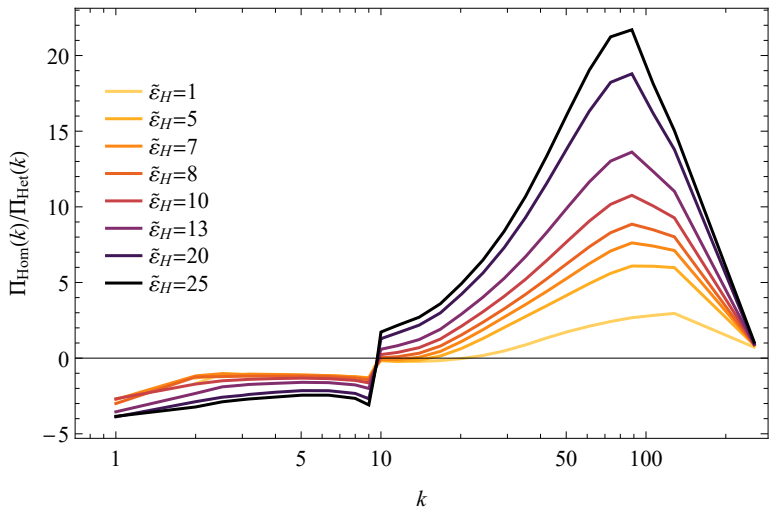}
	\caption{Ratio of the energy flux due to homochiral triads to the energy flux due to heterochiral triads, versus $k$ and for the same parameters as in Fig.~\ref{fig:2}.}
	\label{fig:fluxratio}
\end{figure}

To contrast the strengths of heterochiral and homochiral triadic interactions, we calculate the ratio of their energy fluxes, which is given by
\begin{equation}
\frac{\Pi_{\rm Hom}(k)}{\Pi_{\rm Het}(k)}= \frac{{}^{+}\Pi^{+<}_{+>}(k)+{}^{-}\Pi^{-<}_{->}(k)}{\Pi_E(k)-\left({}^{+}\Pi^{+<}_{+>}(k)+{}^{-}\Pi^{-<}_{->}(k)\right)}.
\end{equation}
In Fig. \ref{fig:fluxratio}, this ratio is plotted versus $k$, for $\tilde{\varepsilon}_H \in S$.
This ratio is negative at scales larger than the energy injection scales, and positive at scales smaller than the energy injection scales. This is in agreement with the dual inverse energy cascade at large scales and forward helicity cascade at small scales. It is also clearly $k$-dependent.  Finally, with the increase of $\tilde{\varepsilon}_H$, the ratio $|\Pi_{\rm Hom}(k)| / |\Pi_{\rm Het}(k)|$ increases without sharp transition.  

\subsection{Non-local energy transfers}
%%%%%%%%%%%%%%%%%
\label{sec:Weakly non local energy transfer}
In order to investigate the degree of locality of the cascades, we further introduce the energy flux from an inner sphere of radius $k_1$ to an outer sphere or radius $k_2$ , which is defined as
\begin{equation}
\Pi^{< k_1}_{> k_2}
= \sum_{|\bk'|\ge k_2}
\Re\{\bu(\bk')^*\cdot {\cal{F}}[(\bu\cdot\nabla)\bu^{<k_1}](\bk')\},
\end{equation}
with
\begin{equation}
\bu^{<k_1}= \sum_{|\bk'|\le k_1} {\cal{F}}^{-1}[\bu(\bk')],
\label{fluxpnl}
\end{equation}
and $k_1 \le k_2$. For $k_1=k_2=k$ the two spheres are adjacent, implying that $\Pi^{< k}_{> k}=\Pi_E(k)$. Increasing the value of $k_2-k_1$ corresponds to moving the two spheres apart, and then to investigate non-local fluxes.\\

\begin{figure}
	\centering
	\includegraphics[scale=0.55]{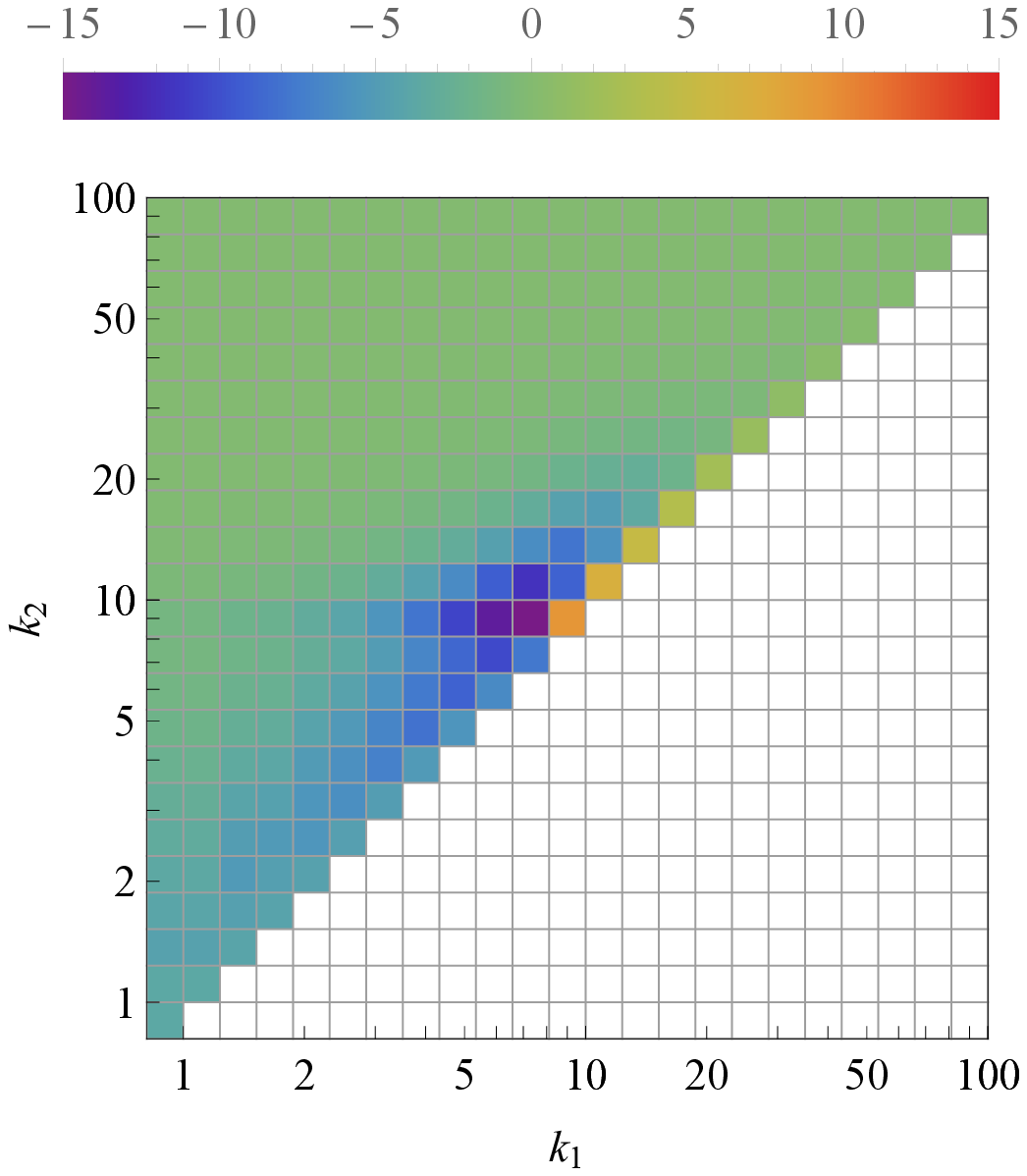}
	\includegraphics[scale=0.55]{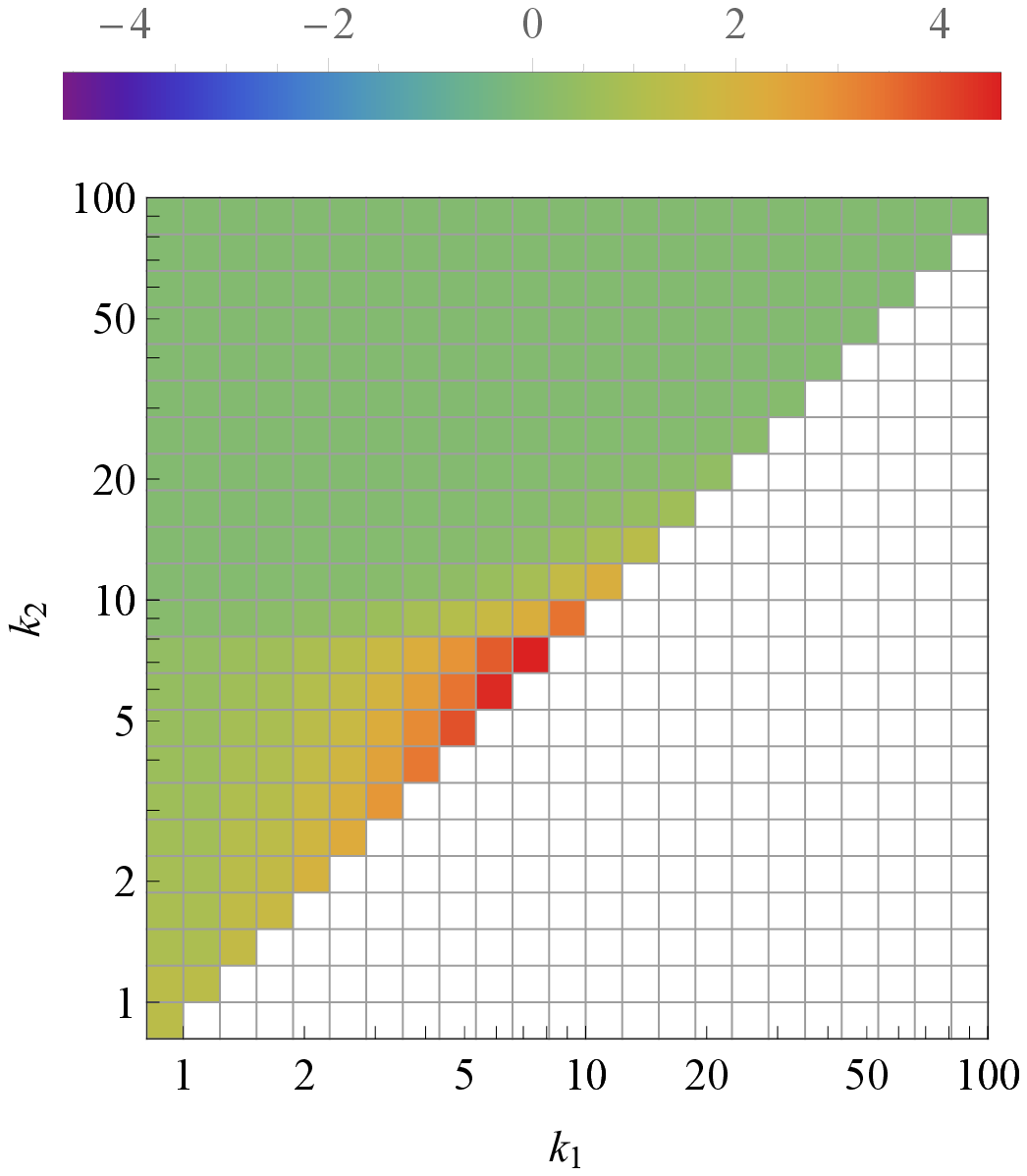}
	\caption{In $(k_1,k_2)$-maps, two representations of fluxes from a $k_1$ inner-sphere to a $k_2$ outer-sphere $\Pi^{< k_1}_{> k_2}$, for $\varepsilon_E=0.2$ and $\tilde{\varepsilon}_H=25$.  In the left map the fluxes are calculated among positive homochiral triads only. In the right map the fluxes are calculated among all other triads, namely heterochiral and negative homochiral. The range of values taken by the fluxes is $[-15;15]$ in left and $[-5;5]$ in right.}
	\label{fig:fluxkk}
\end{figure}

In Fig.~\ref{fig:fluxkk}, two $(k_1,k_2)$ maps of $\Pi^{< k_1}_{> k_2}$ are given, the left map giving the flux among positive homochiral triads, the right map giving the flux among all other triads (heterochiral and negative homochiral). 
In the left map the diagonal coincides with the flux ${}^{+}\Pi^{+<}_{+}(k)$ represented in the right figure of Fig.~\ref{fig:5}. 
In the right map the diagonal coincides with  $\Pi_E(k)-{}^{+}\Pi^{+<}_{+}(k)$.
In both cases the further away from the diagonal, the more non-local the fluxes are. 
In the left map, the fluxes are mostly negative, with a minimum at $(k_1,k_2)\approx (7,9)$.
This shows that the energy cascade is inverse, and due to weakly non-local interactions. 
In the right map the fluxes are positive and maximum on the diagonal, implying a forward cascade due to local interactions. In the right map the fluxes are approximately three times smaller than in the left map, implying that the sum of both is qualitatively similar to the left map. 
Therefore the inverse cascade obtained in the infrared domain is mainly due to weakly non-local interactions among positive homochiral triads $(+,+,+)$. The sum of the other triadic interactions leads to forward cascade, mainly local.
The aforementioned non-local interactions leading to inverse energy cascade are also observed in two-dimensional hydrodynamic turbulence \citep{Gupta2019, Verma2019}.

\subsection{Varying the helicity injection range of scales}
%%%%%%%%%%%%%%%%%
\label{sec:varhel}
In contrast to previous studies \citep{Biferale2012,Biferale2013,Sahoo2017}, here we can vary the range of scales in which helicity is injected. We already know that injecting helicity at scales smaller than the energy injection scales
leads to a forward cascade of energy \citep{Kessar2015}. Here this would correspond to injecting helicity in the range $k_H\in ]9,10^2]$. 
From section \ref{sec:Energy and helicity spectra and fluxes}, we also know that injecting helicity in the range $k_H\in [1,10^2]$ leads to an inverse energy cascade at large scales.

\begin{figure}
	\centering
	\includegraphics[scale=0.75]{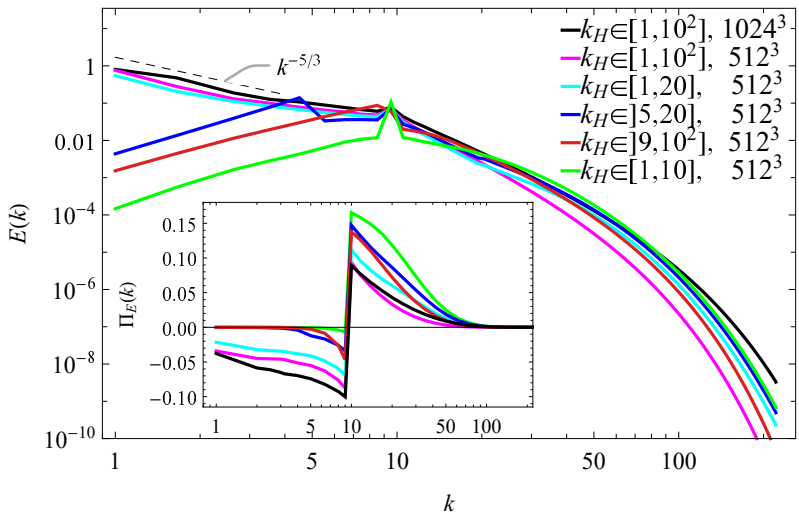}
	\caption{Energy spectra and fluxes (inlet). The energy injection rate $\varepsilon_E=0.2$ is applied at $k_E\in]9,10]$. From bottom to top
of the energy spectra, the helicity injection rate  $\tilde{\varepsilon}_H=25$ is applied at $k_H\in[1,10],
]9,10^2], ]5,20]$, $[1,20]$, $[1,10^2]$ and $[1,10^2]$ again. For the last curve the viscosity is twice smaller and the resolution is equal to $1024^3$.}
	\label{fig:windows}
\end{figure}

We now investigate additional cases injecting helicity in different scale ranges, $k_H\in [1,20],  ]5,20]$ and  $[1,10]$, keeping all other parameters the same. The results of these three cases are plotted in Fig.~\ref{fig:windows}, together with the case $k_H\in [1,10^2]$, already shown in Fig.~\ref{fig:2}, and the case $k_H\in ]9,10^2]$.

The inverse cascade is obtained for $k_H\in [1,20]$ but not for $k_H\in [1,10]$, suggesting that a sufficiently broad range of scales smaller than the energy injection scales is necessary to produce the inverse cascade. This emphasizes the role of non-local transfers identified in section \ref{sec:Weakly non local energy transfer}.
For $k_H\in  ]5,20]$ the energy flux is negative for $k\in  ]5,10]$ with an inverse cascade limited to these scales. This emphasizes the role of helicity injection at scales larger than the energy injection scale, i.e. the inverse energy cascade stops at scales where helicity is not injected.  In summary, to obtain an inverse cascade it is necessary to inject helicity in a sufficiently broad range of scales on both sides of the energy injection range of scales.

Finally, the case $k_H\in [1,10^2]$ with a twice smaller viscosity $\nu=0.75\times10^{-3}$ and a resolution $1024^3$ is 
also plotted in Fig.~\ref{fig:windows}. It corresponds to 
a mean kinetic energy $\bar{E}_{tot}\approx 2.1$, thus leading to $\lambda \approx 0.28$, and a Taylor-microscale Reynolds number $R_{\lambda}\approx 440$.  The energy spectrum and flux are qualitatively similar to the case shown in Fig.~\ref{fig:2}, with again an inverse energy cascade at large scales.

\section{Conclusions}
%%%%%%%%%%%%%%%%%%%%%%%%%%%
The mechanism responsible for the inverse cascade of energy obtained at infrared scales agrees well with the homochiral framework described by \citet{Biferale2012}.
In order to generate a strongly helical turbulence, a positive helicity is injected at all scales, including the infrared scales. We observe a gradual amplification of the inverse cascade with the increase of helicity injection, reaching some saturation at a sufficiently large level of helicity injection. The negative energy flux is due to weakly non-local homochiral triadic interactions. A necessary condition for inverse cascade is to inject helicity in a sufficiently broad range of scales on both sides of the energy injection range of scales. 

We note that this inverse cascade mechanism is different from the one acting in rotating turbulence \citep{Mininni2009} or in a thin layer \citep{Musacchio2019}. Indeed, in 
both these cases, the flow becomes quasi-two-dimensional which is not the case here.
\begin{figure}
	\centering
	\includegraphics[scale=0.825, angle=0]{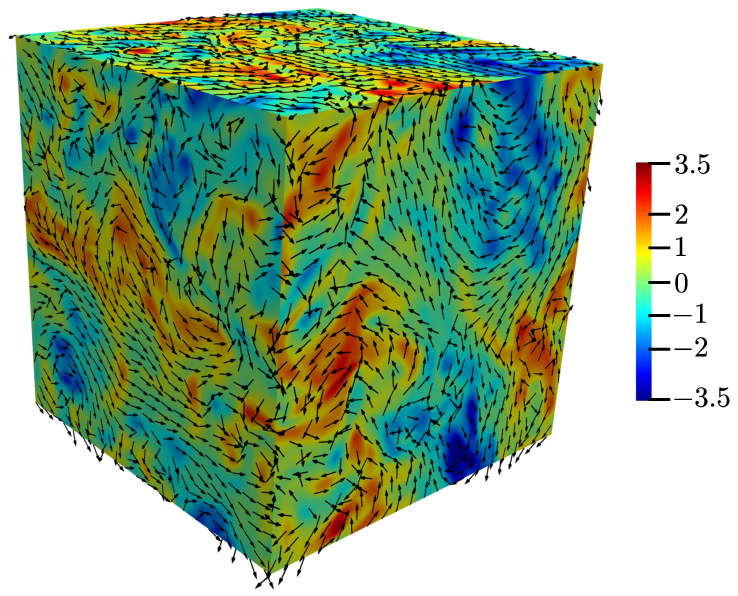}
	\includegraphics[scale=0.825, angle=0]{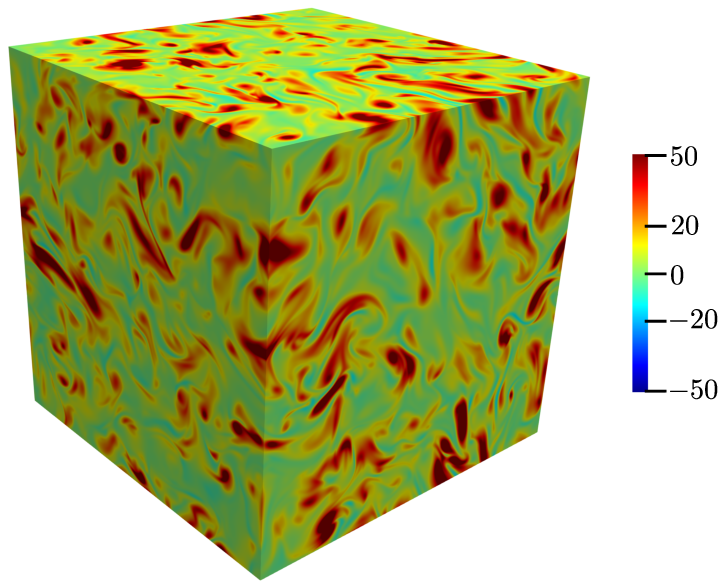}
	\caption{Snapshots of velocity (left) and helicity (right) on the three faces of the cubic resolution domain for $\varepsilon_E=0.2$ and $\tilde{\varepsilon}_H=25$. In the left figure the colors represent the isovalues of the velocity component perpendicular to each face, and the arrows the velocity field parallel to each face. In the right figure the colors represent the isovalues of helicity.}
	\label{fig:cube}
\end{figure}
In Fig.~\ref{fig:cube}, snapshots of both velocity and helicity are represented on the three faces of the resolution domain. We see large-scale structures due to the inverse cascade, but the flow is clearly three-dimensional.

Our results are consistent with the dual cascade phenomenology characterized by an inverse cascade of energy with $E(k)\propto k^{-5/3}$ for $k < k_E$ and a forward cascade of helicity for $k > k_E$ \citep{Alexakis2018}. Finally our study is also consistent with the experimental results of near-maximum helical turbulence by \citet{Herbert2012}, in which a non-local inverse cascade was identified, but for which an alternative explanation based on axisymmetric turbulence was also invoked \citep{Qu2018}.\\
\\
F.P. and M.K.V. are grateful for support from IFCAM project MA/IFCAM/19/90, and CEFIPRA project  6104-1. M.K.V. and R.S. thank Grenoble Alpes University for support.
Our numerical simulations have been performed on Shaheen II at the KAUST supercomputing laboratory, Saudi Arabia, under the project k1416.\\
Declaration of Interests. The authors report no conflict of interest.
\bibliographystyle{jfm}
\bibliography{ref}

\begin{thebibliography}{33}
\expandafter\ifx\csname natexlab\endcsname\relax\def\natexlab#1{#1}\fi
\def\au#1{#1} \def\ed#1{#1} \def\yr#1{#1}\def\at#1{#1}\def\jt#1{\textit{#1}}
  \def\bt#1{#1}\def\bvol#1{\textbf{#1}} \def\vol#1{#1} \def\pg#1{#1}
  \def\publ#1{#1}\def\arxiv#1{#1}\def\org#1{#1}\def\st#1{\textit{#1}}

\bibitem[Alexakis(2017)]{Alexakis2017}
{\sc \au{Alexakis, A.}} \yr{2017}  \at{Helically decomposed turbulence}.
  \jt{J. Fluid Mech.}  \bvol{812},  \pg{752–770}.

\bibitem[Alexakis \& Biferale(2018)]{Alexakis2018}
{\sc \au{Alexakis, A.} \& \au{Biferale, L.}} \yr{2018}  \at{Cascades and
  transitions in turbulent flows}.  \jt{Physics Reports}  \bvol{767-769},
  \pg{1 -- 101}.

\bibitem[{Biferale} {\em et~al.\/}(2012){Biferale}, {Musacchio} \&
  {Toschi}]{Biferale2012}
{\sc \au{{Biferale}, L.}, \au{{Musacchio}, S.} \& \au{{Toschi}, F.}} \yr{2012}
  \at{Inverse energy cascade in three-dimensional isotropic turbulence}.
  \jt{Phys. Rev. Lett.}  \bvol{108}~(16),  \pg{164501}.

\bibitem[{Biferale} {\em et~al.\/}(2013){Biferale}, {Musacchio} \&
  {Toschi}]{Biferale2013}
{\sc \au{{Biferale}, L.}, \au{{Musacchio}, S.} \& \au{{Toschi}, F.}} \yr{2013}
  \at{{Split energy-helicity cascades in three-dimensional homogeneous and
  isotropic turbulence}}.  \jt{J. Fluid Mech.}  \bvol{730},  \pg{309--327}.

\bibitem[{Brissaud} {\em et~al.\/}(1973){Brissaud}, {Frisch}, {Leorat},
  {Lesieur} \& {Mazure}]{Brissaud1973}
{\sc \au{{Brissaud}, A.}, \au{{Frisch}, U.}, \au{{Leorat}, J.}, \au{{Lesieur},
  M.} \& \au{{Mazure}, A.}} \yr{1973}  \at{{Helicity cascades in fully
  developed isotropic turbulence}}.  \jt{Phys. Fluids}  \bvol{16},
  \pg{1366--1367}.

\bibitem[{Cambon} \& {Jacquin}(1989)]{Cambon1989}
{\sc \au{{Cambon}, C.} \& \au{{Jacquin}, J.}} \yr{1989}  \at{{Spectral approach
  to non-isotropic turbulence subjected to rotation}}.  \jt{J. Fluid Mech.}
  \bvol{202},  \pg{295--317}.

\bibitem[Chen {\em et~al.\/}(2003{\natexlab{{\em a\/}}})Chen, Chen \&
  Eyink]{Chen2003}
{\sc \au{Chen, Q.}, \au{Chen, S.} \& \au{Eyink, G.~L.}} \yr{2003{\natexlab{{\em
  a\/}}}}  \at{The joint cascade of energy and helicity in three-dimensional
  turbulence}.  \jt{Phys. Fluids}  \bvol{15}~(2),  \pg{361--374}.

\bibitem[Chen {\em et~al.\/}(2003{\natexlab{{\em b\/}}})Chen, Chen, Eyink \&
  Holm]{Chen2003b}
{\sc \au{Chen, Q.}, \au{Chen, S.}, \au{Eyink, G.~L.} \& \au{Holm, D.~D.}}
  \yr{2003{\natexlab{{\em b\/}}}}  \at{Intermittency in the joint cascade of
  energy and helicity}.  \jt{Phys. Rev. Lett.}  \bvol{90}~(21),  \pg{214503}.

\bibitem[{Craya}(1958)]{Craya1958}
{\sc \au{{Craya}, A.}} \yr{1958}  \at{{Contribution \`a l'analyse de la
  turbulence associ\'ee \`a des vitesses moyennes}}.  \jt{P.S.T. Minist\`ere de
  l'Air (Paris)}  \bvol{345}.

\bibitem[Gupta {\em et~al.\/}(2019)Gupta, Jayaraman, Chatterjee, Sadhukhan,
  Samtaney \& Verma]{Gupta2019}
{\sc \au{Gupta, A.}, \au{Jayaraman, R.}, \au{Chatterjee, A.G.}, \au{Sadhukhan,
  S.}, \au{Samtaney, R.} \& \au{Verma, M.K.}} \yr{2019}  \at{{Energy and
  enstrophy spectra and fluxes for the inertial-dissipation range of
  two-dimensional turbulence}}.  \jt{Phys. Rev. E}  \bvol{100},  \pg{053101}.

\bibitem[{Herbert} {\em et~al.\/}(2012){Herbert}, {Daviaud}, {Dubrulle},
  {Nazarenko} \& {Naso}]{Herbert2012}
{\sc \au{{Herbert}, E.}, \au{{Daviaud}, F.}, \au{{Dubrulle}, B.},
  \au{{Nazarenko}, S.} \& \au{{Naso}, A.}} \yr{2012}  \at{Dual non-{K}olmogorov
  cascades in a von {K}\'{a}rm\'{a}n flow}.  \jt{EPL}  \bvol{100}~(4),
  \pg{44003}.

\bibitem[{Herring}(1974)]{Herring1974}
{\sc \au{{Herring}, J.~R.}} \yr{1974}  \at{{Approach of axisymmetric turbulence
  to isotropy}}.  \jt{Phys. Fluids}  \bvol{17},  \pg{859--872}.

\bibitem[Kessar {\em et~al.\/}(2015)Kessar, Plunian, Stepanov \&
  Balarac]{Kessar2015}
{\sc \au{Kessar, M.}, \au{Plunian, F.}, \au{Stepanov, R.} \& \au{Balarac, G.}}
  \yr{2015}  \at{Non-{K}olmogorov cascade of helicity-driven turbulence}.
  \jt{Phys. Rev. E}  \bvol{92},  \pg{031004}.

\bibitem[Kraichnan(1967)]{Kraichnan1967}
{\sc \au{Kraichnan, R.~H.}} \yr{1967}  \at{Inertial ranges in two‐dimensional
  turbulence}.  \jt{Phys. Fluids}  \bvol{10}~(7),  \pg{1417--1423}.

\bibitem[Lessinnes {\em et~al.\/}(2011)Lessinnes, Plunian, Stepanov \&
  Carati]{Lessinnes2011}
{\sc \au{Lessinnes, T.}, \au{Plunian, F.}, \au{Stepanov, R.} \& \au{Carati,
  D.}} \yr{2011}  \at{Dissipation scales of kinetic helicities in turbulence}.
  \jt{Phys. Fluids}  \bvol{23}~(3),  \pg{035108}.

\bibitem[Mininni {\em et~al.\/}(2006)Mininni, Alexakis \& Pouquet]{Mininni2006}
{\sc \au{Mininni, P.~D.}, \au{Alexakis, A.} \& \au{Pouquet, A.}} \yr{2006}
  \at{Large-scale flow effects, energy transfer, and self-similarity on
  turbulence}.  \jt{Phys. Rev. E}  \bvol{74},  \pg{016303}.

\bibitem[Mininni \& Pouquet(2009)]{Mininni2009}
{\sc \au{Mininni, P.~D.} \& \au{Pouquet, A.}} \yr{2009}  \at{Helicity cascades
  in rotating turbulence}.  \jt{Phys. Rev. E}  \bvol{79},  \pg{026304}.

\bibitem[Moffatt(1969)]{Moffatt1969}
{\sc \au{Moffatt, H.~K.}} \yr{1969}  \at{The degree of knottedness of tangled
  vortex lines}.  \jt{J. Fluid Mech.}  \bvol{35}~(1),  \pg{117–129}.

\bibitem[Moreau(1961)]{Moreau1961}
{\sc \au{Moreau, J.~J.}} \yr{1961}  \at{Constantes d'un \^{i}lot
  tourbillonnaire en fluide parfait barotrope}.  \jt{C. R. Acad. Sci. Paris}
  \bvol{252},  \pg{2810}.

\bibitem[Musacchio \& Boffetta(2019)]{Musacchio2019}
{\sc \au{Musacchio, S.} \& \au{Boffetta, G.}} \yr{2019}  \at{Condensate in
  quasi-two-dimensional turbulence}.  \jt{Phys. Rev. Fluids}  \bvol{4},
  \pg{022602}.

\bibitem[Okamoto {\em et~al.\/}(2007)Okamoto, Yoshimatsu, Schneider, Farge \&
  Kaneda]{Okamoto2007}
{\sc \au{Okamoto, N.}, \au{Yoshimatsu, K.}, \au{Schneider, K.}, \au{Farge, M.}
  \& \au{Kaneda, Y.}} \yr{2007}  \at{Coherent vortices in high resolution
  direct numerical simulation of homogeneous isotropic turbulence: A wavelet
  viewpoint}.  \jt{Phys. Fluids}  \bvol{19}~(11),  \pg{115109}.

\bibitem[Plunian {\em et~al.\/}(2019)Plunian, Stepanov \& Verma]{Plunian2019}
{\sc \au{Plunian, F.}, \au{Stepanov, R.} \& \au{Verma, M.K.}} \yr{2019}  \at{On
  uniqueness of transfer rates in magnetohydrodynamic turbulence}.  \jt{J.
  Plasma Phys.}  \bvol{85}~(5),  \pg{905850507}.

\bibitem[Pope(2000)]{Pope2000}
{\sc \au{Pope, S.~B.}} \yr{2000} {\em Turbulent flows\/}.  \publ{Cambridge
  Univ. Press}.

\bibitem[Qu {\em et~al.\/}(2018)Qu, Naso \& Bos]{Qu2018}
{\sc \au{Qu, B.}, \au{Naso, A.} \& \au{Bos, W. J.~T.}} \yr{2018}  \at{Cascades
  of energy and helicity in axisymmetric turbulence}.  \jt{Phys. Rev. Fluids}
  \bvol{3},  \pg{014607}.

\bibitem[Sadhukhan {\em et~al.\/}(2019)Sadhukhan, Samuel, Plunian, Stepanov,
  Samtaney \& Verma]{Sadhukhan2019}
{\sc \au{Sadhukhan, S.}, \au{Samuel, R.}, \au{Plunian, F.}, \au{Stepanov, R.},
  \au{Samtaney, R.} \& \au{Verma, M.K.}} \yr{2019}  \at{Enstrophy transfers in
  helical turbulence}.  \jt{Phys. Rev. Fluids}  \bvol{4},  \pg{084607}.

\bibitem[Sahoo {\em et~al.\/}(2017)Sahoo, Alexakis \& Biferale]{Sahoo2017}
{\sc \au{Sahoo, G.}, \au{Alexakis, A.} \& \au{Biferale, L.}} \yr{2017}
  \at{Discontinuous transition from direct to inverse cascade in
  three-dimensional turbulence}.  \jt{Phys. Rev. Lett.}  \bvol{118},
  \pg{164501}.

\bibitem[{Stepanov} {\em et~al.\/}(2015){Stepanov}, {Golbraikh}, {Frick} \&
  {Shestakov}]{2015PhRvL}
{\sc \au{{Stepanov}, R.}, \au{{Golbraikh}, E.}, \au{{Frick}, P.} \&
  \au{{Shestakov}, A.}} \yr{2015}  \at{Hindered energy cascade in highly
  helical isotropic turbulence}.  \jt{Phys. Rev. Lett.}  \bvol{115}~(23),
  \pg{234501}.

\bibitem[{Stepanov} {\em et~al.\/}(2018){Stepanov}, {Teimurazov}, {Titov},
  {Verma}, {Barman}, {Kumar} \& {Plunian}]{Stepanov2017}
{\sc \au{{Stepanov}, R.}, \au{{Teimurazov}, A.}, \au{{Titov}, V.}, \au{{Verma},
  M.~K.}, \au{{Barman}, S.}, \au{{Kumar}, A.} \& \au{{Plunian}, F.}} \yr{2018}
  \at{Direct numerical simulation of helical magnetohydrodynamic turbulence
  with {T}arang code}.  \jt{2017 Ivannikov ISPRAS Open Conference (ISPRAS)}
  \pg{pp. 90--96. IEEE}.

\bibitem[{Teimurazov} {\em et~al.\/}(2018){Teimurazov}, {Stepanov}, {Verma},
  {Barman}, {Kumar} \& {Sadhukhan}]{Teimurazov2018}
{\sc \au{{Teimurazov}, A.~S.}, \au{{Stepanov}, R.~A.}, \au{{Verma}, M.~K.},
  \au{{Barman}, S.}, \au{{Kumar}, A.} \& \au{{Sadhukhan}, S.}} \yr{2018}
  \at{{Direct Numerical Simulation of Homogeneous Isotropic Helical Turbulence
  with the TARANG Code}}.  \jt{J. Appl. Mech. Tech. Phys.}  \bvol{59}~(7),
  \pg{1279--1287}.

\bibitem[Verma(2019)]{Verma2019}
{\sc \au{Verma, M.K.}} \yr{2019} {\em Energy Transfers in Fluid Flows:
  Multiscale and Spectral Perspective\/}.  \publ{Cambridge Univ. Press}.

\bibitem[{Verma}(2004)]{Verma2004}
{\sc \au{{Verma}, M.~K.}} \yr{2004}  \at{{Statistical theory of
  magnetohydrodynamic turbulence: recent results}}.  \jt{Phys. Rep.}
  \bvol{401},  \pg{229--380}.

\bibitem[{Verma} {\em et~al.\/}(2013){Verma}, Chatterjee, Yadav, Paul, Chandra
  \& Samtaney]{Verma2013}
{\sc \au{{Verma}, M.~K.}, \au{Chatterjee, A.~G.}, \au{Yadav, R.~K.}, \au{Paul,
  S.}, \au{Chandra, M.} \& \au{Samtaney, R.}} \yr{2013}  \at{{Benchmarking and
  scaling studies of pseudospectral code Tarang for turbulence simulations}}.
  \jt{Pramana-J. Phys.}  \bvol{81},  \pg{617–629}.

\bibitem[Waleffe(1992)]{Waleffe1992}
{\sc \au{Waleffe, F.}} \yr{1992}  \at{The nature of triad interactions in
  homogeneous turbulence}.  \jt{Phys. Fluids}  \bvol{4}~(2),  \pg{350--363}.

\end{thebibliography}

\end{document}